\documentclass{PoS}

\title{Probing technicolor theories with staggered fermions }

\ShortTitle{Probing technicolor theories with staggered fermions }

\author{Zolt\'{a}n Fodor \\
Department of Physics, University of Wuppertal, Gauss
  Strasse 20, D-42119, Germany \\
E-mail: \email{fodor@bodri.elte.hu}}

\author{\speaker{Kieran Holland} \\
Department of Physics, University of the Pacific\\
  3601 Pacific Ave, Stockton CA 95211, USA \\
E-mail: \email{kholland@pacific.edu}}

\author{Julius Kuti \\
  Department of Physics 0319, University of California, San Diego \\
  9500 Gilman Drive, La Jolla CA 92093, USA \\
E-mail: \email{jkuti@ucsd.edu}}

\author{D\'{a}niel N\'{o}gr\'{a}di \\
  Department of Physics 0319, University of California, San Diego \\
  9500 Gilman Drive, La Jolla CA 92093, USA \\
E-mail: \email{nogradi@lorentz.leidenuniv.nl}}

\author{Chris Schroeder \\
  Department of Physics 0319, University of California, San Diego \\
  9500 Gilman Drive, La Jolla CA 92093, USA \\
E-mail: \email{crs@physics.ucsd.edu}}

\abstract{One exciting possibility of new physics beyond the Standard
  Model is that the fundamental Higgs sector is replaced by a
  strongly-interacting gauge theory, known as technicolor. A viable
  theory must break chiral symmetry dynamically, like in QCD, to
  generate Goldstone bosons which become the longitudinal components
  of the $W^{\pm}$ and $Z$. By measuring the eigenvalues of the Dirac
  operator, one can determine if chiral symmetry is in fact spontaneously
  broken. We simulate $SU(3)$ gauge theory with $n_s=2$ and 3 staggered
  flavors in the fundamental representation, corresponding to $N_f=8$
  and 12 flavors in the continuum limit. 
  Although our first findings show that both theories are consistent
  with dynamically broken chiral symmetry and QCD-like behavior,
  flavor breaking effects in the spectrum may require further
  clarifications before final conclusions can be drawn.
  We also compare various improved staggered actions, to
  suppress this potentially large flavor breaking.
  } 

\FullConference{The XXVI International Symposium on Lattice Field Theory \\
		 July 14 - 19, 2008\\
		 Williamsburg, Virginia, USA}

\begin{document}

\section{Introduction}

The LHC will probe the mechanism of electroweak symmetry breaking. A
very attractive alternative to the standard Higgs mechanism, with
fundamental scalars, involves new strongly-interacting gauge theories,
known as technicolor \cite{Weinberg:1979bn, Susskind:1978ms}. Such models
avoid difficulties of theories with scalars, such as triviality and
fine-tuning. Chiral symmetry must be spontaneously broken in a technicolor
theory, to provide the technipions which generate the $W^{\pm}$ and
$Z$ masses and break electroweak symmetry. Although this duplication
of QCD is appealing, precise electroweak measurements have made it
difficult to find a viable candidate theory. It is also necessary to
enlarge the theory (extended technicolor) to generate quark masses,
without generating large flavor-changing neutral currents, which is
challenging. 

\begin{figure}[t]
\begin{center}
\includegraphics[width=.4\textwidth]{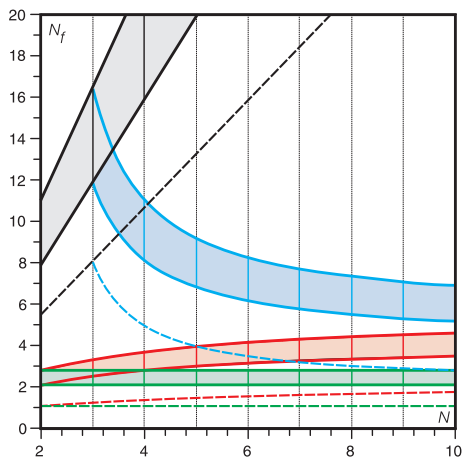}
\end{center}
\caption{The conformal window for $SU(N)$ gauge theories with $N_f$
  techniquarks in various representations, from
  \cite{Dietrich:2006cm}. The shaded regions are the windows, for
  fundamental (gray), 2-index antisymmetric (blue), 2-index symmetric
  (red) and adjoint (green) representations.} 
\label{fig:window}
\end{figure}

Technicolor theories have lately enjoyed a resurgence, due to the
exploration of various techniquark representations
\cite{Dietrich:2006cm}. Feasible candidates have fewer new flavors,
reducing tension with electroweak constraints. If a theory is almost
conformal, it is possible this generates additional energy scales,
which could help in building the extended technicolor sector. There
are estimates of which theories are conformal for various
representations, shown in Fig.~\ref{fig:window}. For $SU(N)$ gauge
theory, if the number of techniquark flavors is less than some
critical number, conformal and chiral symmetries are broken and the
theory is QCD-like. For future model-building, it is crucial to go
beyond these estimates and determine precisely where the conformal windows
are. There have been a number of recent lattice simulations of
technicolor theories, attempting to locate the conformal windows for
various representations \cite{Catterall:2007yx, Appelquist:2007hu,
  Shamir:2008pb, Deuzeman:2008sc, DelDebbio:2008zf}. 

\section{Dirac eigenvalues and chiral symmetry}

\begin{figure}[t]
\begin{center}
\includegraphics*[width=.7\textwidth]{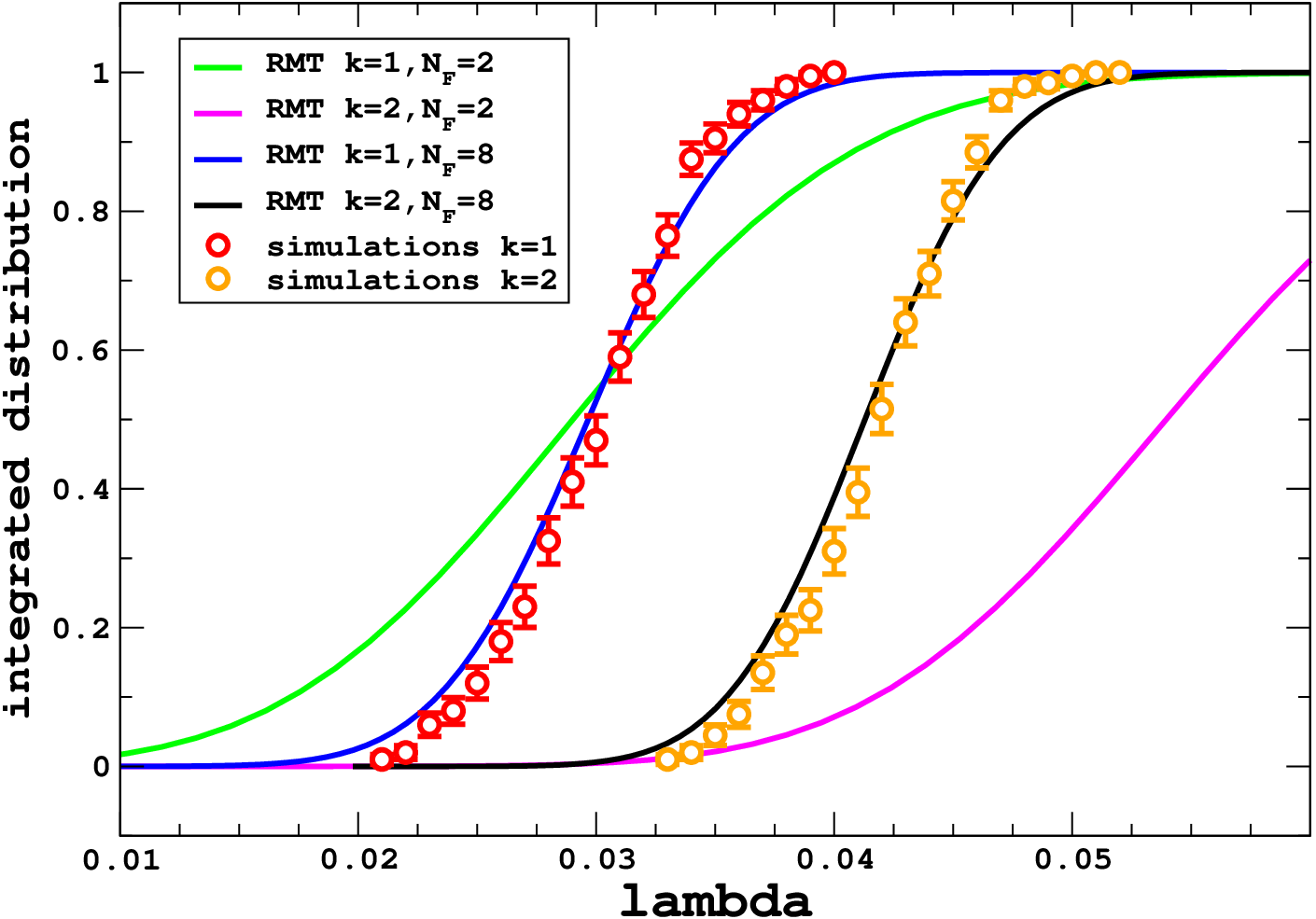}
\end{center}
\caption{The integrated distribution of the two lowest eigenvalue
  quartets, from simulations of $n_s=2$ Asqtad staggered flavors. This
  is compared to RMT with $N_f=2$ and 8, corresponding to the strong
  and weak coupling limits.}
\label{fig:2stag}
\end{figure}

The connection between the eigenvalues $\lambda$ of the Dirac operator and
chiral symmetry breaking is succinctly given in the Banks-Casher
relation \cite{Banks:1979yr},
\begin{equation}
\Sigma = - \langle \bar{\Psi} \Psi \rangle = \lim_{\lambda \rightarrow
  0} \lim_{m \rightarrow 0} \lim_{V \rightarrow \infty} \frac{\pi
  \rho(\lambda)}{V}. 
\label{eq:Banks}
\end{equation}
To generate a non-zero density $\rho(0)$, the smallest eigenvalues
must become densely packed as the volume increases, with an
eigenvalue spacing $\Delta \lambda \approx 1/\rho(0) = \pi/(\Sigma
V)$. This allows a crude estimate of the quark condensate
$\Sigma$. One can do much better by exploring the $\epsilon$-regime:
If chiral symmetry is spontaneously broken, tune the volume and
quark mass such that
\begin{equation}
\frac{1}{F_\pi} \ll L \ll \frac{1}{m_\pi},
\label{eq:epsilon}
\end{equation}
so that the pion is much lighter than the physical value, and
finite-volume effects are dominant \cite{Gasser:1987ah}. The chiral
Lagrangian,  
\begin{equation}
{\cal L} = \frac{F_\pi^2}{4} {\rm Tr}( \partial_\mu U \partial_\mu
U^\dagger ) + \frac{\Sigma}{2} {\rm Tr} [ M(U + U^\dagger) ],
\hspace{0.5cm} U = \exp \left[ \frac{i \pi^a T^a}{F_\pi} \right]
\label{eq:chiral}
\end{equation}
is dominated by the zero-momentum mode from the mass term and all
kinetic terms are suppressed. In this limit, the distributions of the
lowest eigenvalues are identical to those of random matrix theory
(RMT), a theory of large matrices obeying certain symmetries
\cite{Shuryak:1992pi}. To connect with RMT, the eigenvalues and quark
mass are rescaled as $z = \lambda \Sigma V$ and $\mu = m \Sigma V$,
and the eigenvalue distributions also depend on the topological charge $\nu$
and the number of quark flavors $N_f$. RMT is a very useful tool to
calculate analytically all of the eigenvalue distributions. The
eigenvalue distributions in various topological sectors are measured
via lattice simulations, and via comparison with RMT, the value of the
condensate $\Sigma$ can be extracted. This method has been
successfully used in a number of lattice QCD studies, for example in
dynamical overlap fermion simulations \cite{Fukaya:2007yv}.  

\section{Simulations and analysis}

For $SU(3)$ gauge theory with quarks in the fundamental representation,
various methods suggest that the critical number of flavors
separating conformal and QCD-like behavior is between 8 and 12. In
order to study this interesting region, we simulate $n_s=2$ and 3
staggered fermion flavors, corresponding to $N_f=8$ and 12 flavors in the
continuum limit. (We do not take roots of the determinant of the
staggered Dirac operator). We have also simulated $SU(3)$ gauge theory
with $N_f=2$ flavors in the 2-index symmetric representation, using
dynamical overlap fermions, which is described in \cite{Fodor:2S}. We
use the Asqtad staggered action \cite{Orginos:1999cr}, which includes
improvements to reduce the violations of flavor symmetry (``taste
breaking'') at finite lattice spacing. This action is very well tested
and has been heavily used in large scale simulations of lattice QCD
\cite{Davies:2003ik}. There have been detailed comparisons of
staggered eigenvalues with the Asqtad action to RMT
\cite{Follana:2005km}, but only in the quenched approximation.    

\begin{figure}[t]
\begin{center}
\includegraphics*[width=.7\textwidth]{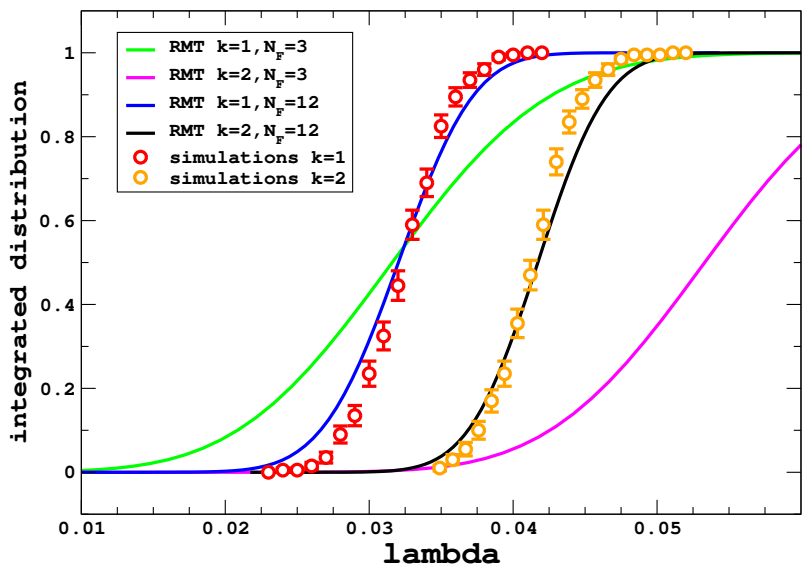}
\end{center}
\caption{The integrated distribution of the two lowest eigenvalue
  quartets, from simulations of $n_s=3$ Asqtad staggered flavors. This
  is compared to RMT with $N_f=3$ and 12, corresponding to the strong
  and weak coupling limits.}
\label{fig:3stag}
\end{figure}

Because $n_s=2$ and 3 staggered flavors have not been simulated with
this action before, a large scan of the parameter space of the bare
couplings was required. Hence our first runs were on small volumes
$10^4$, where we also gained experience on the dependence of the
Hybrid Monte Carlo algorithm \cite{Gottlieb:1987mq} on the quark mass
and the discretization of the trajectory length. Once we generated
large thermalized ensembles, we calculated the lowest eigenvalues of the
Dirac operator using the PRIMME package \cite{primme}. In the
continuum limit, the staggered eigenvalues form degenerate quartets,
with restored flavor symmetry. In Figs.~\ref{fig:2stag} and
\ref{fig:3stag}, we show the integrated distributions of the two
lowest eigenvalue quartet averages,
\begin{equation}
\int_0^{\lambda} p_k(\lambda') d\lambda', \hspace{0.5cm} k=1,2
\end{equation} 
for ensembles with $n_s=2$ and 3 staggered flavors respectively. Both
simulations have quark mass $ma=0.01$, and the respective bare
couplings are $\beta=3.9$ and 1.9. All low eigenvalues have
small chirality, with no indication of non-zero topology. We see that
the quark mass is less than the average smallest eigenvalue, which is
necessary to probe the behavior of the eigenvalue distributions in the
chiral limit. To compare with RMT, we vary $\mu=m \Sigma V$ until we
satisfy 
\begin{equation}
\frac{\langle \lambda_1 \rangle_{\rm sim}}{m} = \frac{\langle z
  \rangle_{\rm rmt}}{\mu},
\label{eq:rmt}
\end{equation}
where $\langle \lambda_1 \rangle_{\rm sim}$ is the lowest quartet average
from simulations and the RMT average $\langle z \rangle_{\rm rmt}$
depends implicitly on $\mu$ and $N_f$. With this optimal value of
$\mu$, we can predict the distributions $p_k(\lambda')$ and compare to
the simulations.   

\begin{figure}[t]
\begin{center}
\includegraphics*[width=.7\textwidth]{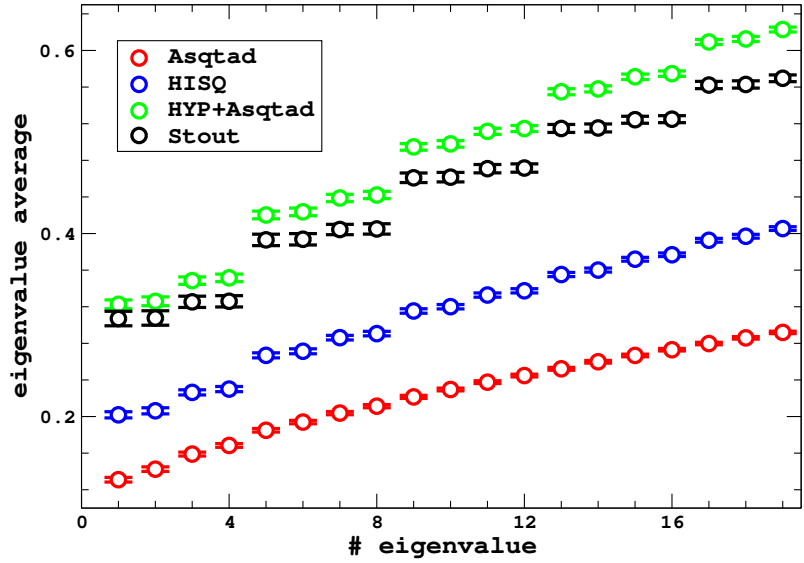}
\end{center}
\caption{Comparison of different improvements of the staggered Dirac
  operator. The eigenvalues are calculated on the same ensemble of
  gauge configurations, which were generated using the Asqtad action.}
\label{fig:impstag}
\end{figure}

In both cases, we see quite good agreement between simulations and RMT
with the corresponding number of flavors in the continuum limit
i.e.~$N_f=8$ and 12. This is somewhat surprising. From the eigenvalues
themselves, one can directly see that flavor breaking is significant,
since degenerate quartets are not yet formed. A previous
eigenvalue study used unimproved staggered quarks in dynamical fermion
simulations \cite{Damgaard:2000qt}. They found excellent agreement
with RMT but only if $N_f$ had the same value as the number of
staggered flavors $n_s$. We also find that, at strong coupling, RMT
with the continuum value $N_f=4 n_s$ does not describe the data. On coarse
lattices, the flavor breaking is very large and only one pion can be
tuned to the $\epsilon$-regime for each staggered flavor. One has to
go to weak coupling and finer lattices, where flavor breaking
decreases, to recover the correct number of light pions.  

\begin{figure}[t]
\begin{center}
\includegraphics*[width=.7\textwidth]{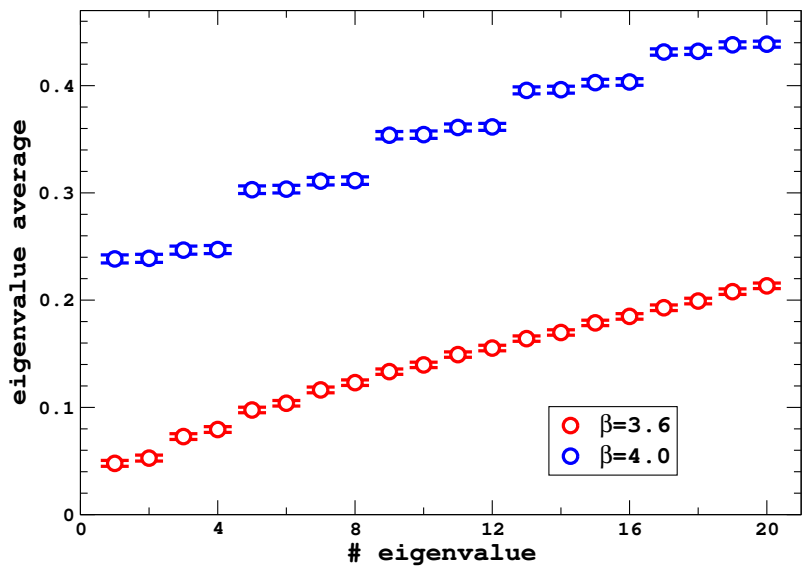}
\end{center}
\caption{The lowest eigenvalues calculated on two ensembles with
  $n_s=1$ staggered flavor, with stout smearing used both in the sea
  and valence quark. The lattice volume is $12^4$.}
\label{fig:stout}
\end{figure}

These results indicate that both the $N_f=8$ and 12 flavor
theories with fundamental quarks have a non-zero quark condensate
$\Sigma$ i.e.~chiral symmetry is spontaneously broken. 
If this conclusion holds against further studies of flavor breaking
effects, our $N_f = 8$ result will lend considerable support to the
findings of \cite{Appelquist:2007hu, Deuzeman:2008sc}, but the $N_f = 12$
spectrum would be inconsistent with the statement of
\cite{Appelquist:2007hu} that this theory is conformal.

\section{Staggered improvement}

Since flavor breaking can have a dramatic effect on the eigenvalues, we
are investigating various improvements of the staggered action, to
bring the simulations closer to the continuum limit. In
Fig.~\ref{fig:impstag}, we compare mixed actions, with gauge
configurations generated using the Asqtad action, while the
eigenvalues are those of various improved staggered Dirac
operators. This figure is for $n_s=1$ staggered flavor at $\beta=6.8$
and volume $10^4$. The appearance of eigenvalue quartets which are
clearly separated is a clear indication of reduced flavor
breaking. Both HYP-smearing \cite{Hasenfratz:2001hp} and
stout-smearing \cite{Morningstar:2003gk} seem to bring significant
improvement relative to the Asqtad operator, while HISQ fermions
\cite{Follana:2006rc} do not show as clear an improvement. 

We also show in Fig.~\ref{fig:stout} the effect of using
stout-smearing both in the sea and valence quark. As we go to weaker
coupling towards the continuum limit, the eigenvalue quartet structure
emerges clearly. Comparison of the improved eigenvalues with RMT is
ongoing. 

\section{Conclusions}

Knowledge of the conformal window is essential to build viable
candidates of strongly interacting physics beyond the Standard
Model, and lattice simulations will play a crucial role. Our technique of
studying the eigenvalue properties complements other lattice
approaches, such as calculating the beta function of the renormalized
coupling, looking for finite-temperature transitions, or extracting
the mass spectrum. This will hopefully lead to consensus about the
nature of these new theories. Our first study gives an indication that
$SU(3)$ gauge theory with $N_f=8$ and 12 flavors are both QCD-like,
non-conformal theories. We are investigating various improvements to
reduce flavor-breaking lattice artifacts and allow us to reach a stronger
conclusion. 

\section{Acknowledgments}
We thank Poul Damgaard for very helpful discussions, and Urs Heller
who stressed the importance of reaching the quartet degeneracy
limit. This research was supported by the DOE under grants
DOE-FG03-97ER40546, DE-FG02-97ER25308, by the NSF under grant 0704171,
by the DFG under grant FO 502/1 and by SFB-TR/55.

\end{document}